\documentclass[a4paper,11pt]{article}
\usepackage{pos}
\usepackage{float}
\usepackage{caption}
\usepackage{subcaption}
\usepackage{enumitem}
\usepackage{setspace}
\usepackage{lineno}

\title{Status of the MicroBooNE Low Energy Excess Search}

\manuallySeparateAuthors
\author[a]{David Caratelli}
\author{, for the MicroBooNE Collaboration}

\affiliation[a]{Fermilab, PO Box 500, Batavia, IL, 60510, USA}

\emailAdd{davidc@fnal.gov}

\abstract{MicroBooNE is a neutrino experiment that utilizes a liquid argon time projection chamber (LArTPC) located on-axis in the Booster Neutrino Beam (BNB) at Fermilab. One of the experiment’s main goals is to search for excess low-energy electromagnetic-like events as seen by the MiniBooNE experiment, located just downstream of MicroBooNE in the BNB. As MicroBooNE nears the completion of its first single-electron-like and single-photon-like searches, these proceedings present the status of MicroBooNE's low-energy excess search as of early summer 2020. In addition to presenting an overview of the approach to these analyses, we showcase results from $\pi^0$ calibrations and $e$/$\gamma$ separation, and sample results from sidebands aimed at validating the analysis' progress outside the low-energy signal region.}

\FullConference{%
  40th International Conference on High Energy physics - ICHEP2020\\
  July 28 - August 6, 2020\\
  Prague, Czech Republic (virtual meeting)
}

\begin{document}
\maketitle

\section{Introduction}
The MicroBooNE detector~\cite{ubdetector} was built to follow-up on the observation of an excess of low-energy electromagnetic (EM) activity by the MiniBooNE collaboration~\cite{miniboonelee}. After five years of data-taking, MicroBooNE has collected approximately 12E20 protons on target (POT) of data from the Booster Neutrino Beamline (BNB) and has made significant progress towards producing first results in its follow-up study of the MiniBooNE low-energy excess. These proceedings present an overview of the progress in calibration, pattern-recognition, neutrino interaction modeling and other aspects vital to the analysis, and presents preliminary results from sideband data aimed at validating the analysis' performance before final box-opening. 

\subsection{Electron and Photon Low-Energy Excess Searches}

\begin{figure}[H] 
\begin{center}
    \begin{subfigure}[b]{0.4\textwidth}
    \centering
    \includegraphics[width=1.00\textwidth]{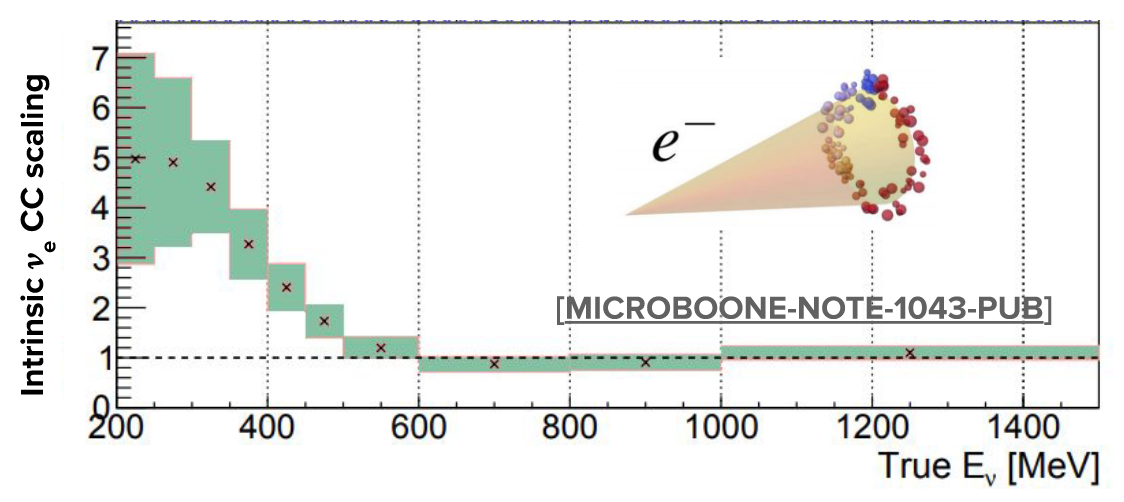}
    \caption{\label{fig:escaling} $\nu_e$ scaling}
    \end{subfigure}
    \begin{subfigure}[b]{0.4\textwidth}
    \centering
    \includegraphics[width=1.00\textwidth]{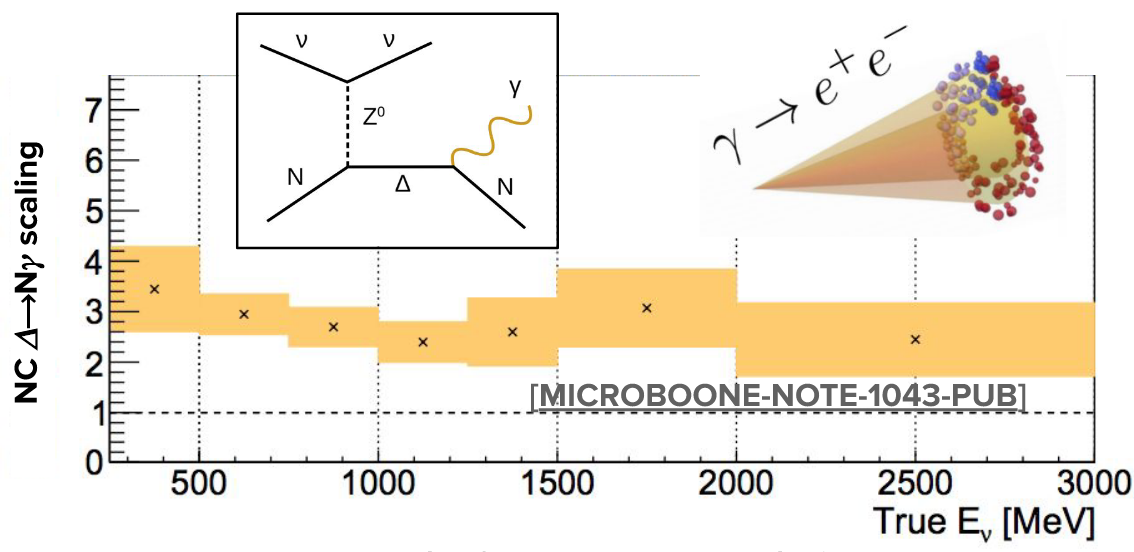}
    \caption{\label{fig:gscaling} NC $\Delta \rightarrow \gamma$ scaling}
    \end{subfigure}
\caption{\label{fig:leemodels} Model for a $\nu_e$ (left) and NC $\Delta$-radiative low-energy excess signal constructed from MiniBooNE's excess observation, from MicroBooNE's \href{https://microboone.fnal.gov/wp-content/uploads/MICROBOONE-NOTE-1043-PUB.pdf}{PUBLIC-NOTE-1043}.}
\end{center}
\end{figure}

\paragraph{}MicroBooNE's primary experimental goal is to investigate the nature of a possible source of low-energy EM events and determine whether they are associated to electrons or photons. To meet this goal, MicroBooNE has active analyses exploring both hypotheses. Currently one photon-like search and multiple electron-like searches are being pursued. In their current state, MicroBooNE's electron and photon searches target specific signal definitions, both tied to neutrino interactions leading to single EM shower final-states. Electron searches target the measurement of $\nu_e$ interactions at low energy, while the photon search is aiming to measure the rate of NC $\Delta$-resonant production followed by radiative single-photon decay. In order to benchmark the analysis' performance, a data-driven signal model is constructed starting from MiniBooNE's excess data observation for both the $\nu_e$ and NC $\Delta$-radiative hypotheses, providing an energy-dependent scaling of intrinsic $\nu_e$ and a flat scaling of NC $\Delta$-radiative interactions necessary to account for the MiniBooNE observation in MicroBooNE, shown in figure~\ref{fig:leemodels}. 

As the first large-scale LArTPC in the US-based accelerator-neutrino program, MicroBooNE has been a breeding ground for the development of novel reconstruction and analysis methods which aim to leverage the high-quality images of neutrino interactions which this technology provides. In keeping to its R\&D goals, MicroBooNE is utilizing multiple reconstruction paradigms in its effort to investigate the MiniBooNE low-energy excess. Specifically, three separate analyses are being conducted in the electron search. Each analysis approach and the tools it leverages is briefly summarized below:
\begin{itemize}[noitemsep]
    \item \textbf{1$e$1$p$ Deep Learning based search} Employing novel deep-learning based tools~\cite{dl,ssnet,mpid} this analysis aims to measure $\nu_e$ interactions with a proton and electron in the final-state with kinematics consistent with the quasi-elastic (QE) interaction process. The status of this analysis is documented in MicroBooNE's \href{https://microboone.fnal.gov/wp-content/uploads/MICROBOONE-NOTE-1086-PUB.pdf}{PUBLIC-NOTE-1086}.
    \item \textbf{Pandora $\nu_e$ search} Relying on the Pandora multi-algorithm reconstruction tool-kit~\cite{pandora} this analysis aims to measure the orthogonal 1$e$N$p$0$\pi$ and 1$e$0$p$0$\pi$ (with N $\geq 1$ and zero charged or neutral pions) channels leveraging calorimetric and topological features in order to perform a kinematics-agnostic measurement of $\nu_e$ interactions in the BNB. The status of this analysis is documented in MicroBooNE's \href{https://microboone.fnal.gov/wp-content/uploads/MICROBOONE-NOTE-1085-PUB.pdf}{PUBLIC-NOTE-1085}.
    \item \textbf{inclusive Wire-Cell $\nu_e$  search} This analysis effort leverages the Wire-Cell 3D imaging reconstruction approach~\cite{wc} with the aim of performing an inclusive measurement of $\nu_e$ interactions. The status of this analysis is documented in MicroBooNE's \href{https://microboone.fnal.gov/wp-content/uploads/MICROBOONE-NOTE-1088-PUB.pdf}{PUBLIC-NOTE-1088}.
\end{itemize}

The \textbf{single-photon search} relies on the Pandora pattern-recognition to measure events with a single photon and one or no protons in the final-state which are consistent with the kinematics of a NC $\Delta$ radiative decay interaction. The status of this analysis is documented in MicroBooNE's \href{https://microboone.fnal.gov/wp-content/uploads/MICROBOONE-NOTE-1087-PUB.pdf}{PUBLIC-NOTE-1087}.

\subsection{Low-Energy Excess Search Status}

\begin{figure}[h] 
\begin{center}
    \begin{subfigure}[b]{0.33\textwidth}
    \centering
    \includegraphics[width=1.00\textwidth]{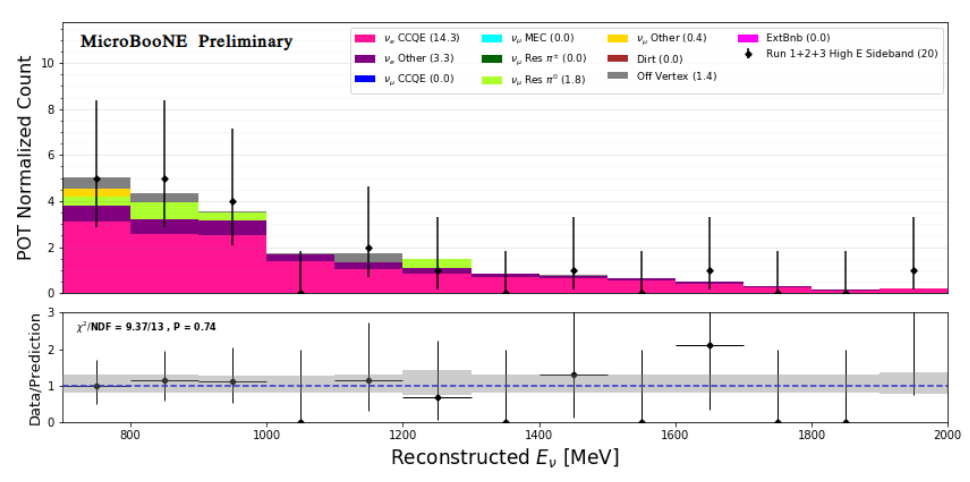}
    \caption{\label{fig:pi0mass:all} {\tiny DL 1$e$1$p$  sideband $E_{\rm reco} > 0.7$ GeV.} }
    \end{subfigure}
    \begin{subfigure}[b]{0.31\textwidth}
    \centering
    \includegraphics[width=1.00\textwidth]{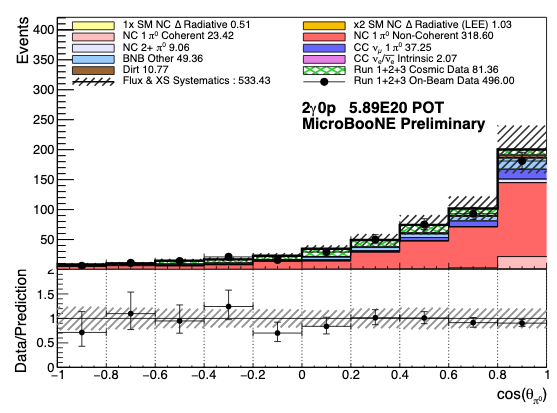}
    \caption{\label{fig:pi0massnc} {\tiny 2$\gamma$0$p$ NC sideband.} }
    \end{subfigure}
    \begin{subfigure}[b]{0.28\textwidth}
    \centering
    \includegraphics[width=1.00\textwidth]{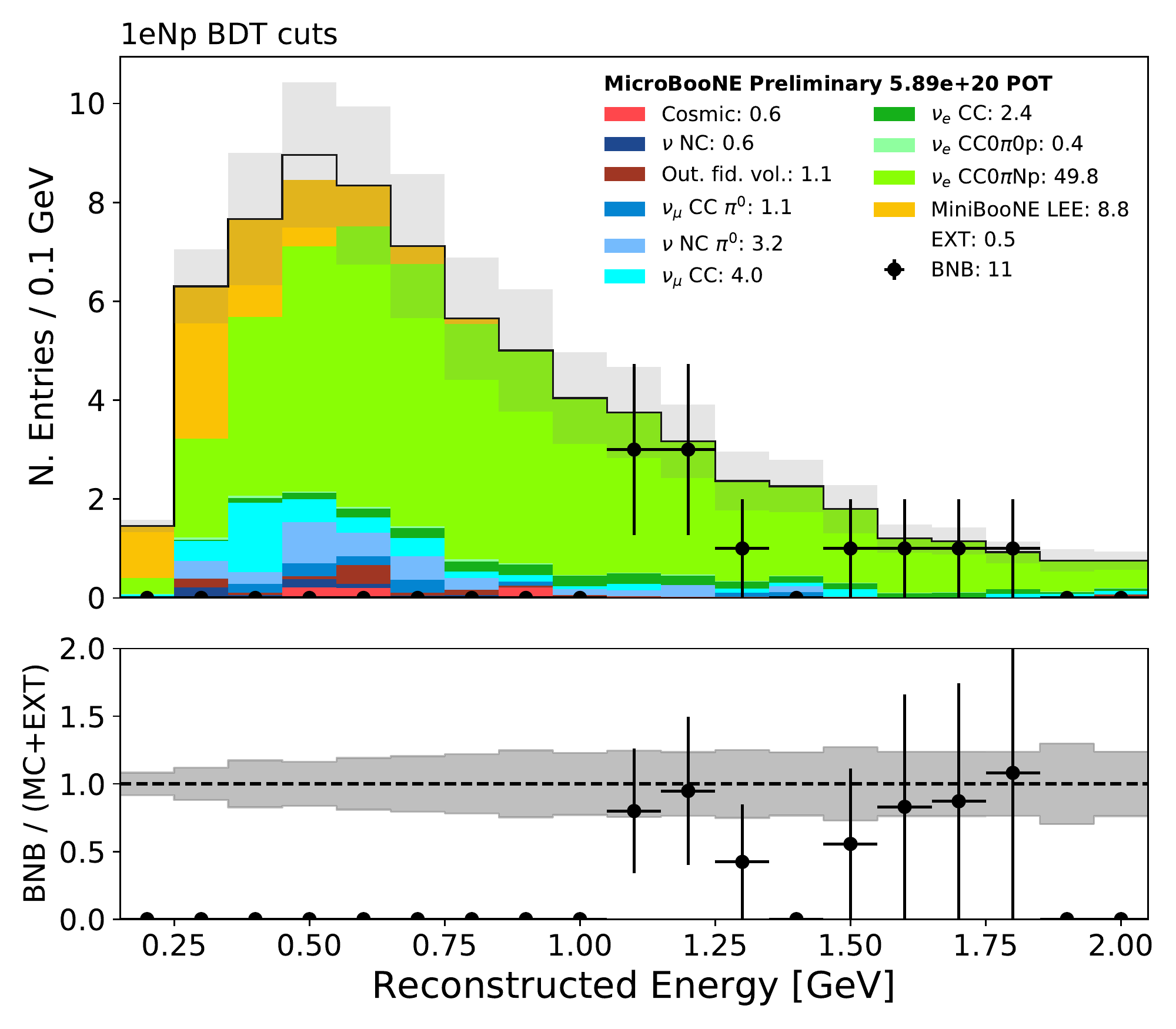}
    \caption{\label{fig:pi0massnc} {\tiny Pandora 1$e$N$p$ sideband $E_{\rm reco} > 1.05$ GeV.} }
    \end{subfigure}
\caption{\label{fig:sidebands} Select sidebands from the Deep Learning based analysis (left), single-photon analysis (center) and Pandora electron analysis (right). Higher resolution images of these and other sideband results can be found in MicroBooNE Public notes~\href{https://microboone.fnal.gov/wp-content/uploads/MICROBOONE-NOTE-1086-PUB.pdf}{1086},~\href{https://microboone.fnal.gov/wp-content/uploads/MICROBOONE-NOTE-1087-PUB.pdf}{1087}, and~\href{https://microboone.fnal.gov/wp-content/uploads/MICROBOONE-NOTE-1085-PUB.pdf}{1085} respectively.}
\end{center}
\end{figure}

\paragraph{}MicroBooNE is in the process of completing its first round of low-energy excess analyses in the photon and electron channels. Ongoing analyses utilize 6.86E20 POT of collected data from the first three years of operation, roughly half the total collected POT by the experiment. The analysis is a blind analysis, with access to 0.6E20 POT of unbiased open data for analysis development. While at different stages of development, these analyses are demonstrating mature tools and performance capable of shedding light on the nature of the origin of MiniBooNE's anomalous low-energy excess. Five years of active development on detector modeling, calibrations, and reconstruction techniques, as well as neutrino interaction modeling and in-situ cross-section measurements have made the experiment a mature environment for precision neutrino measurements that meet the requirements needed for this analysis to succeed. Several of the ongoing analyses have performed in-depth studies of high-statistics $\nu_{\mu}$ interactions and $\nu_e$ sidebands at high-energy and are preparing for final unblinding. Examples of sideband distributions are shown in figure~\ref{fig:sidebands} from the high-energy $\nu_e$ sidebands of the deep-learning and Pandora based $\nu_e$ searches and the NC $\pi^0$ sideband of the single-photon search. The good level of agreement between data and simulation give confidence in the analysis' status and performance. 

\section{Analysis Approach and Tools}

This section briefly describes the general features of MicroBooNE's low-energy excess analysis. All neutrino analyses in MicroBooNE rely on cosmic-rejection techniques to isolate neutrino interactions. This is achieved through the use of scintillation light for triggering~\cite{ublight} and the employment of charge-to-light matching techniques to further isolate neutrino interaction candidates~\cite{wc}. Combined to additional analysis-specific cuts, these tools lead to a factor of $\mathcal{O}\left(10^{8}\right)$ reduction in cosmic-rays, with a rate of $\mathcal{O}$(0-1) expected cosmic-ray backgrounds at the final selection stage.

\begin{figure}[H] 
\begin{center}
    \begin{subfigure}[b]{0.43\textwidth}
    \centering
    \includegraphics[width=1.00\textwidth]{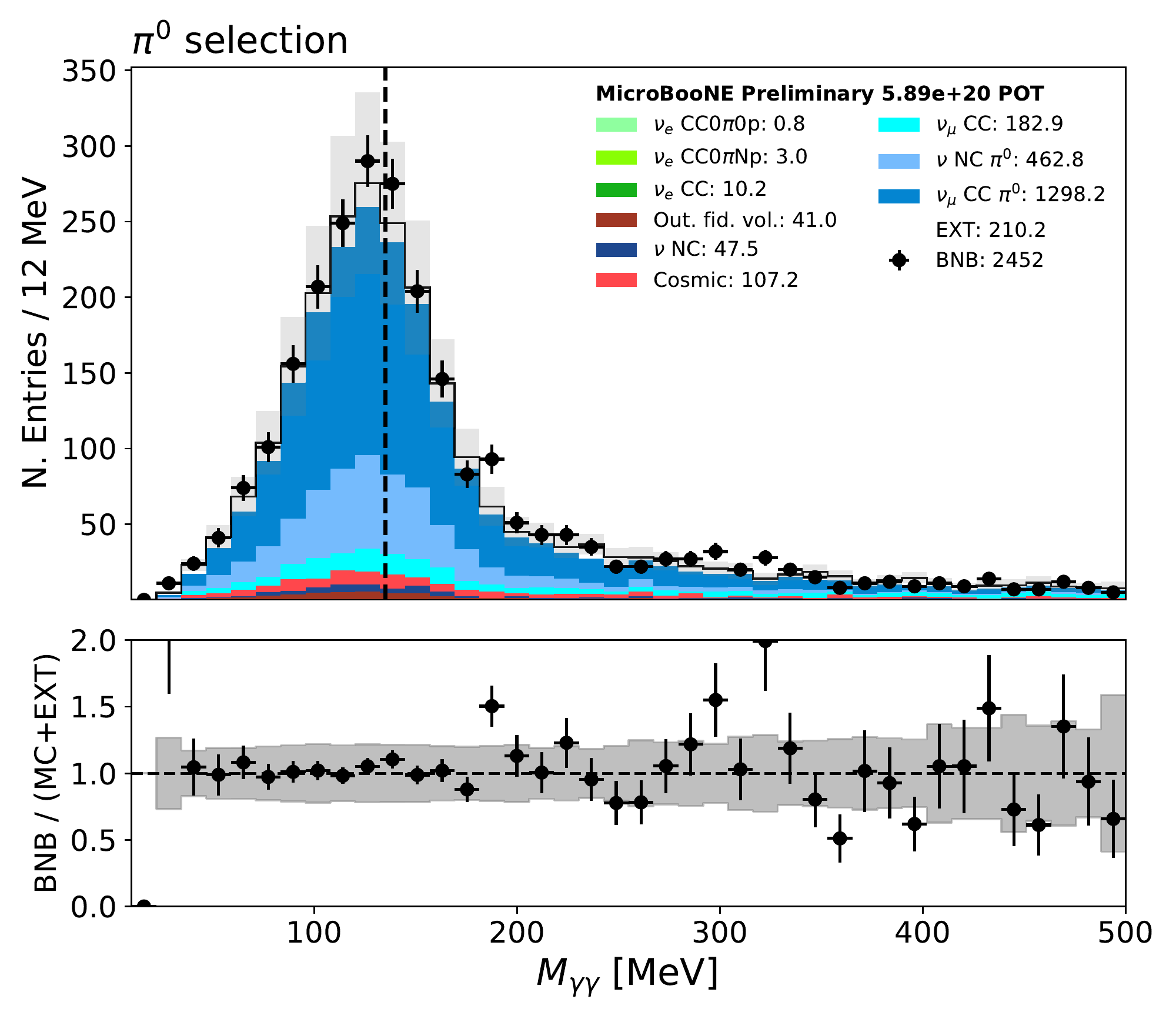}
    \caption{\label{fig:pi0mass:all} NC and CC $\pi^0$ candidates.}
    \end{subfigure}
    \begin{subfigure}[b]{0.52\textwidth}
    \centering
    \includegraphics[width=1.00\textwidth]{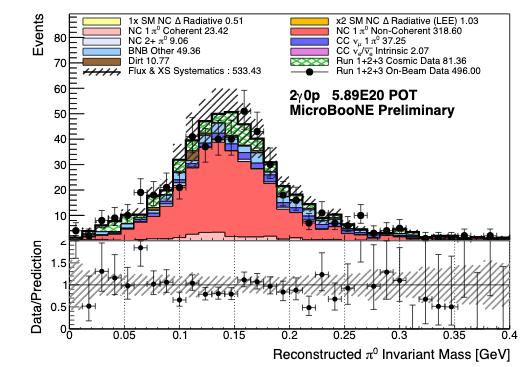}
    \caption{\label{fig:pi0massnc} NC $\pi^0$ candidates.}
    \end{subfigure}
\caption{\label{fig:pi0mass} Reconstructed $M_{\gamma\gamma}$ invariant mass from $\pi^0$ candidates. Left: area-normalized distribution from NC and CC candidates. Right: absolutely normalized distribution from the NC $\pi^0$ measurement.}
\end{center}
\end{figure}

Detector modeling and calibrations are crucial to a well understood background and signal prediction. MicroBooNE's low-energy excess analyses rely on multiple years of iterative detector response and calibration development~\cite{sp1,sp2,efield,calib,pi0}, which have enabled accurate kinematics measurements and powerful particle identification tools. Due to the nature of the analysis, reconstruction and calibrations of EM showers are particularly critical. Figure~\ref{fig:pi0mass} show high statistics distributions of the $\pi^0$ mass from reconstructed $\gamma\gamma$ candidates providing a very good data-MC agreement (relying on simulation-based energy corrections) and absolute energy-scale validation of EM showers at better than 5\%.

\begin{figure}[H] 
\begin{center}
    \begin{subfigure}[b]{0.48\textwidth}
    \centering
    \includegraphics[width=1.00\textwidth]{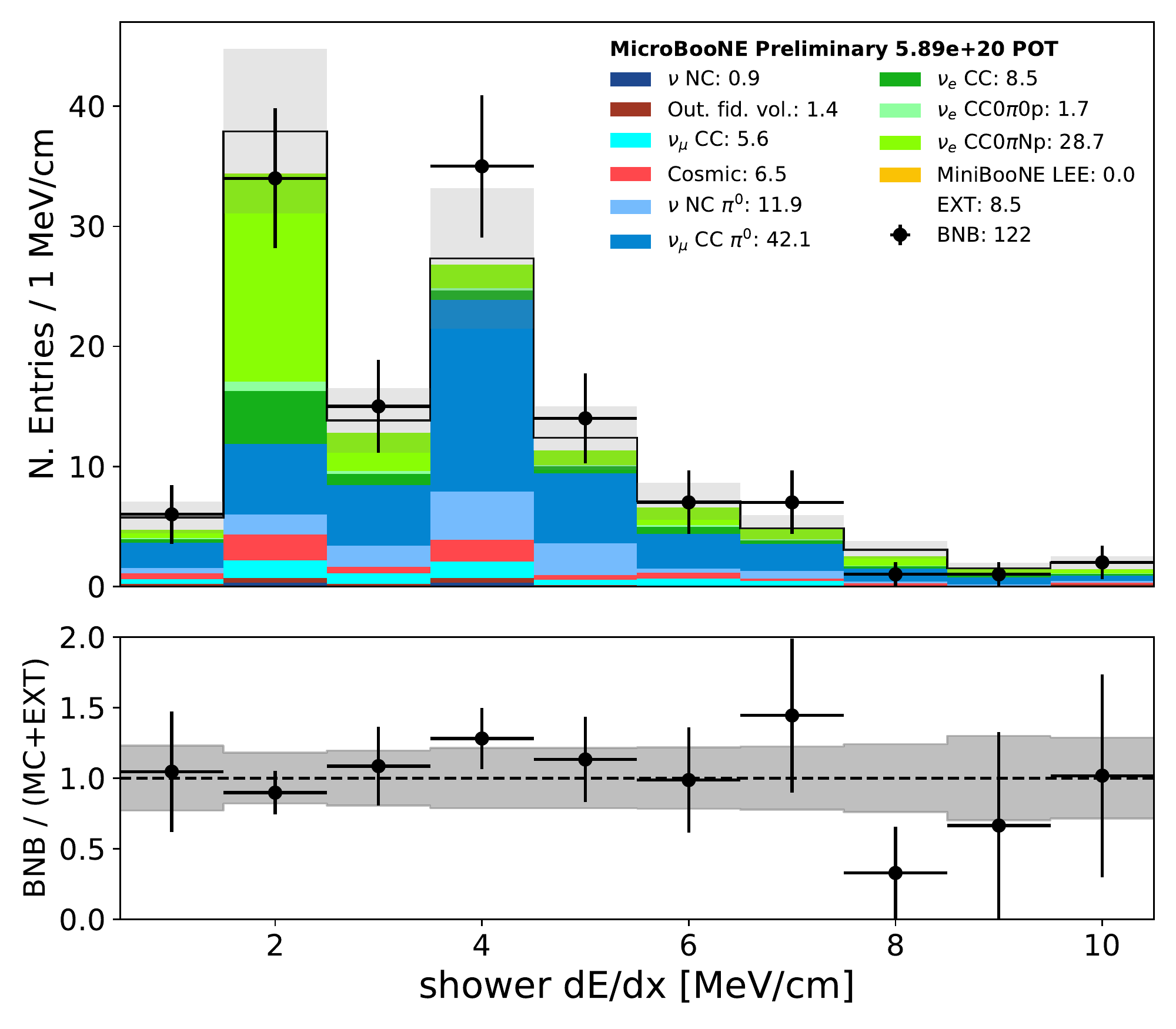}
    \caption{\label{fig:dedx:np} 1$e$N$p$0$\pi$.}
    \end{subfigure}
    \begin{subfigure}[b]{0.48\textwidth}
    \centering
    \includegraphics[width=1.00\textwidth]{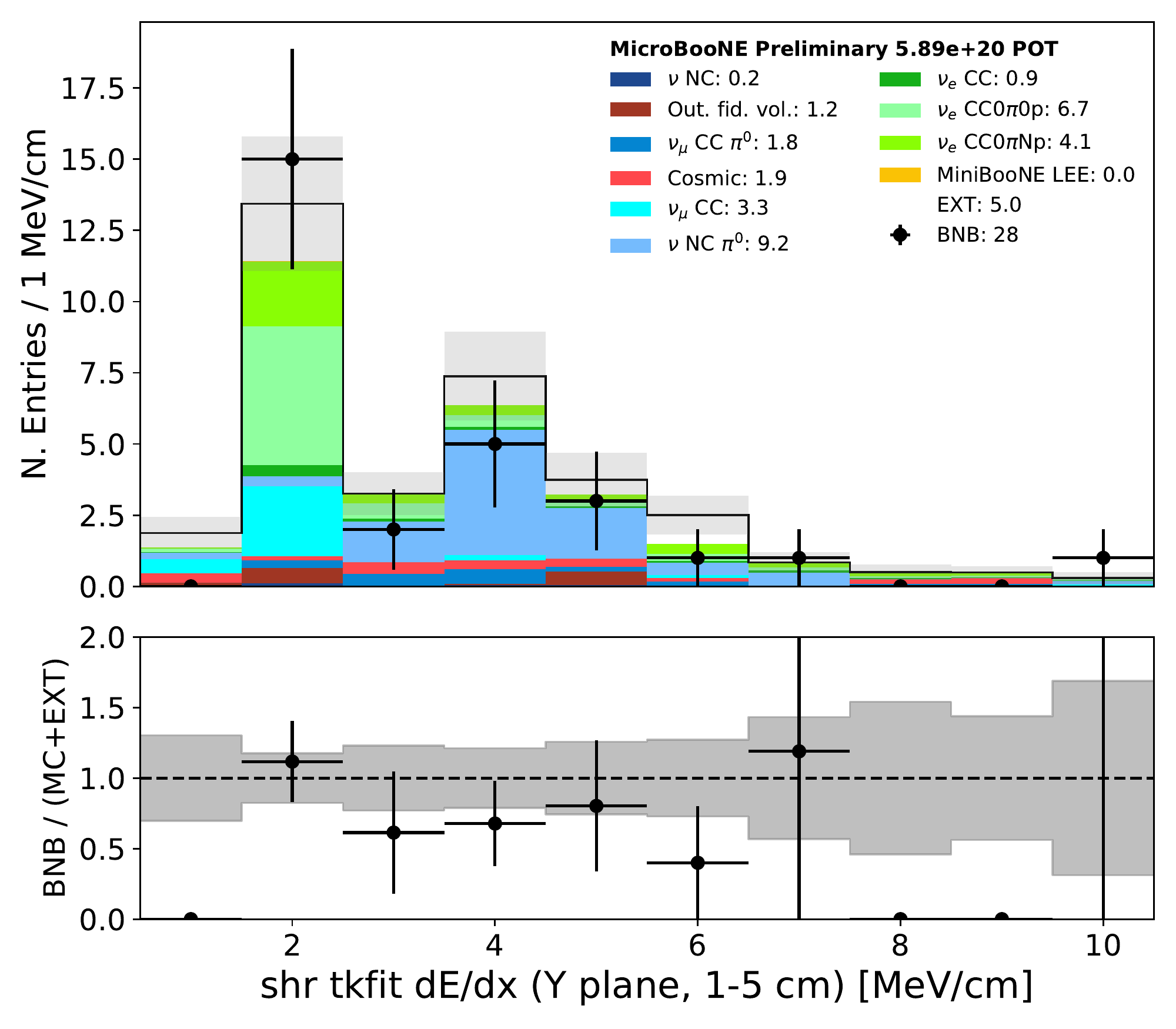}
    \caption{\label{fig:dedx:0p} 1$e$0$p$0$\pi$.}
    \end{subfigure}
\caption{\label{fig:dedx} dE/dx on high energy electron neutrinos after intermediate selection cuts. The left plot comes from the 1$e$N$p$ selection high-energy sideband ($E_{\rm reco} > 1.05$ GeV). The plot on the right is from the 1$e$0$p$ selection high-energy sideband ($E_{\rm reco} > 0.9$ GeV). Sidebands were defined in order to allow adequate $\nu_e$ statistics while remaining blind to potential signal events at low energy.}
\end{center}
\end{figure}

Particle identification, and in particular $e$/$\gamma$ separation, is another critical task for this analysis. High-energy $\nu_e$ sidebands are used to demonstrate and validate the use of the calorimetric $e$/$\gamma$ separation power which the LArTPC technology provides and which is paramount to achieving high-purity measurements of $\nu_e$ and single-$\gamma$ interactions in MicroBooNE's $\nu_{\mu}$ dominated neutrino beam. Figure~\ref{fig:dedx} shows the d$E$/d$x$ measured along the trunk of EM showers at an intermediate selection stage in the 1$e$N$p$0$\pi$ (\ref{fig:dedx:np}) and 1$e$0$p$0$\pi$ (\ref{fig:dedx:0p}) channels from the Pandora based $\nu_e$ analysis.

\begin{figure}[h]
    \centering
    \includegraphics[width=0.5\textwidth]{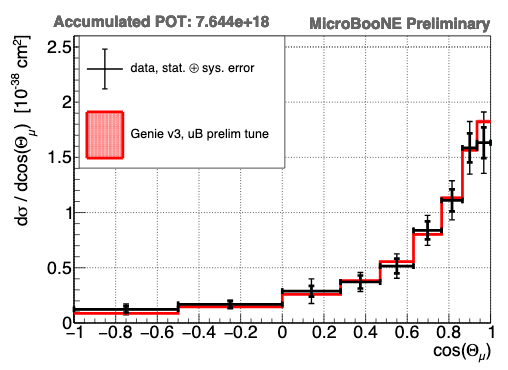}
    \caption{Selected $\nu_{\mu}$ CC interactions with a limited (2\%) dataset. Result from MicroBooNE's \href{https://microboone.fnal.gov/wp-content/uploads/MICROBOONE-NOTE-1069-PUB.pdf}{PUBLIC-NOTE-1069}.}
    \label{fig:numutheta}
\end{figure}

An accurate modeling of neutrino interactions in the detector is necessary in order to properly account for $\nu_{\mu}$ backgrounds and for accurate predictions of the expected rates of intrinsic $\nu_e$ and NC $\pi^0$ interactions. 
Ongoing MicroBooNE analyses rely on the GENIE event generator~\cite{genie}, utilizing the GENIE 3.0.6  G18\_10a\_02\_11a model, further tuned with external CC$0\pi$ data from the T2K collaboration~\cite{t2k} as described in MicroBooNE's \href{https://microboone.fnal.gov/wp-content/uploads/MICROBOONE-NOTE-1074-PUB.pdf}{PUBLIC-NOTE-1074}. The good degree of agreement between data and simulation, as seen in the preliminary comparison in $\nu_{\mu}$ CC lepton $\theta$ shown in figure~\ref{fig:numutheta}, indicates significant progress within the experiment in obtaining a reliable model of neutrino interactions on argon, achieving the maturity necessary for the precision measurements that the low-energy-excess search analysis demands.

Systematic uncertainties on background events drive the sensitivity to potential new physics. Flux modeling uncertainties leverage the work of the MiniBooNE collaboration~\cite{miniboonelee}. Neutrino interaction uncertainties are treated following the implementation within the GENIE model employed, as described in \href{https://microboone.fnal.gov/wp-content/uploads/MICROBOONE-NOTE-1074-PUB.pdf}{PUBLIC-NOTE-1074}. Detector uncertainties rely on data-driven measurements of detector response in a broad range of observable parameters. Their treatment is described in MicroBooNE's \href{https://microboone.fnal.gov/wp-content/uploads/MICROBOONE-NOTE-1075-PUB.pdf}{PUBLIC-NOTE-1075}. To mitigate the impact of modeling uncertainties in the electron and photon channels, each analysis relies on sideband measurements for a data-driven systematic constraint. Leveraging commonalities in the flux parentage and interaction model of electron and muon neutrinos, systematic uncertainties for intrinsic $\nu_e$ interactions are constrained via high-statistics measurements of $\nu_{\mu}$ events. Likewise, measurements of NC $\pi^0$ interactions help constrain uncertainties for the single-photon search. By leveraging the strong correlations between these different measurements, the analyses are able to achieve a significant reduction in systematic uncertainties, enabling measurements with sensitivity to new physics.

\section{Outlook}
The MicroBooNE collaboration is currently completing the last stages of unblinding for its first low energy excess analysis results. The work presented here builds on multiple years of ground-breaking developments in simulation, calibrations, reconstruction and data-analysis techniques for LArTPC detectors, and leverages the experience of MicroBooNE's own cross-section program as well as the broader neutrino interaction community's work on interaction modeling and generator development. The results of MicroBooNE's low-energy searches will provide long-awaited insight into one of the most outstanding recent anomalies in the neutrino sector, and serve as a valuable stepping stone for the broader Short Baseline Neutrino program to be carried out at Fermilab. 

\begingroup
\setstretch{0}

\endgroup
\end{document}